\documentclass[12pt]{article}
\usepackage{amssymb}

\newcommand{\preprintno}[1]
{\vspace{-2cm}{\normalsize\begin{flushright}#1\end{flushright}}\vspace{1cm}}

\title{\preprintno{{\bf MCTP-00-07}}
CP violation and target-space modular invariance}
\author{Thomas Dent\thanks{email: tdent@umich.edu} \\
	{\em Michigan Center for Theoretical Physics} \\
        {\em Randall Laboratory, Department of Physics,} \\
        {\em University of Michigan, Ann Arbor, MI 48109 U.\ S.\ A.}}
\date{November 2000}

\begin{document}

\maketitle

\begin{abstract}
We show that, in perturbative string models where the source of CP violation is a complex vacuum expectation value (v.e.v.)\ for one or more compactification moduli, CP is conserved if a CP transformation acting on the modulus values is an element of a target-space \mbox{(self-)duality} group. Where the duality group is SL$(2,\mathbb{Z})$ the result confirms a conjecture of Bailin {\em et al.\/}\ that CP is conserved for v.e.v.'s of the $T$ modulus on the boundary of the fundamental domain, and generalises Giedt's result on the removability of complex Yukawa couplings in such models. Our result applies to any model of spontaneous CP violation where the CP-odd scalar transforms under a symmetry that is not explicitly broken. We consider whether similar results could be obtained in ``brane worlds''.
\end{abstract}

\section{Introduction: CP violation in perturbative string theory} 
The origin of CP violation in particle physics is an unsolved problem that should eventually be addressed by any prospective ``theory of everything'' - in particular string theory and its recent extensions to ``M-theory'' and D-brane models. Within the range of string theories, the origin of CP violation has received much attention only for the perturbative heterotic string. Strominger and Witten \cite{StroWitten85} established that a suitable extension of the four-dimensional CP operator should reverse the orientation of three (real) compactified dimensions, which is equivalent to complex conjugating the three complex dimensions $Z^\alpha$ of the Calabi-Yau manifold or orbifold used in compactifying the heterotic string. This transformation takes matter representations $R$ to their complex conjugates $\overline{R}$ and allows for CP to be broken by the geometry of the compactification manifold, if it is not symmetric under $Z^\alpha\mapsto Z^{\alpha*}$. 

It was then shown by Choi {\em et al.\/}\ \cite{ChoiKN} that CP can be a {\em gauge symmetry}\/ in a class of theories with $d$ extra dimensions compactified {\em \`a la}\/ Kaluza-Klein, for $d=5,6,7\mbox{ mod }8$ and certain choices of gauge group in $(4+d)$ dimensions. This class then includes the (effective field theory of the) ten-dimensional heterotic string with gauge group E$_8\otimes\mbox{E}_8$ or SO$(32)$; Dine {\em et al.\/}\ \cite{DineLM} obtained an equivalent result working directly with the string theory. The CP transformation constructed by these authors includes (in addition to the usual parity transformation) an orientation-changing Lorentz transformation acting on the compactified directions, a general coordinate transformation also acting on the compactified directions which reverses all Kaluza-Klein gauge quantum numbers corresponding to isometries of the compact space, and an inner automorphism of the $(4+d)$-dimensional gauge group which takes each state in a representation to its complex conjugate.

Since gauge symmetries cannot be explicitly broken \cite{Krauss:1989zc}, CP must be violated {\em spontaneously}\/ in such theories, either by the construction of the compactification or by scalar v.e.v.'s in the low-energy effective field theory. Calabi-Yau manifolds defined by complex parameters can break CP \cite{StroWitten85}, however the origin of such parameters is unknown, particularly given the CP-conserving nature of string dynamics, and concrete models are absent. Kobayashi and Lim \cite{KobayashiL} showed that orbifold compactifications (strictly, $\mathbb{Z}_N$ orbifolds decomposable into the product of three two-dimensional orbifolds) allow a CP transformation of the type described, so leave CP intact. We will consider only spontaneous CP violation in the effective field theory.

Non-zero imaginary parts for the compactification moduli $T^\alpha$, representing the v.e.v.\ of the background antisymmetric tensor field in the compactified directions, are natural candidates for CP-violating quantities, being odd under the orientation-changing transformation of the compact directions. This source of spontaneous CP violation would feed through to the Yukawa couplings (see {\em e.g.\/}\ \cite{BailinLS93}) and soft supersymmetry-breaking terms of the low-energy effective theory \cite{IbanezLust,Bailin:1998iz+97}. The imaginary part of the dilaton $S$ can also act as a source of CP violation since its v.e.v.\ corresponds to the (tree-level) theta angle of the QCD vacuum, and depending on the mechanism of supersymmetry-breaking a nonzero imaginary part for $S$ or for the auxiliary field $F^S$ could produce CP-violating soft terms. Concrete predictions for CP violation depend on being able to calculate the modulus-dependence of the Yukawa couplings and on the mechanism for breaking supersymmetry and stabilizing the moduli and dilaton.

In a series of papers Bailin {\em et al.\/}\ \cite{Bailin:1998iz+97,Bailin:1998xx} investigated the stabilization of moduli at complex values in heterotic string models, using gaugino condensation in the presence of $T$-dependent threshold corrections, and showed that complex Yukawa couplings and soft terms could arise. However, the interpretation of their results is somewhat unclear, since the assignment of the Standard Model fields to the charged matter sectors is not specified, and in fact the models appear to have unrealistic gauge group and matter spectrum. While additional symmetry-breaking mechanisms could solve the latter problems, they are likely to complicate the calculation of CP violating couplings ({\em e.g.\/}\ \cite{Bailin:1998yt}). We might hope that the results of simplified models of the origin of CP violation would have some features in common with a more realistic solution. However, as we will show, such simplified models are constrained by the target-space duality invariance that is generic in heterotic string models.

In any model which generates complex coupling constants, there is the possibility that the complex phases may be removed by redefining the basis of matter fields, implying that the phases are unphysical and do not cause CP violation. This was demonstrated by Giedt \cite{Giedt2000} for a particular set of complex Yukawa couplings calculated in ref.\ \cite{Bailin:1998xx}, under the assumption that particular twisted sector states could be identified with the chiral superfields of the MSSM. 

In this paper we show, using the duality symmetry acting on the moduli and observable fields, that a particular class of modulus v.e.v.'s are CP-conserving, {\em i.e.\/}\ that they lead to complex phases which are in all cases unphysical. Our result is model-independent, in that it can be applied to all models of spontaneous CP violation in which the relevant v.e.v.'s transforms under some other (spontaneously broken) symmetry. The reasoning also avoids the uncertainty which occurs when coupling constants are predicted which are modulus-dependent but {\em not}\/ modular invariant, due to the transformation properties of the matter fields (see below). In heterotic string models with an SL$(2,\mathbb{Z})$ invariance acting on the $T$ modulus, we show that CP is conserved by values of $T$ on the boundary of the fundamental domain $\mathcal{F}$, a conjecture previously made \cite{BKL_unpub} based on the results of various explicit calculations \cite{Bailin:1998iz+97}. We also consider the possibility that CP may be violated in softly-broken supersymmetry by the v.e.v.\ of the dilaton even while $T$-dependent couplings conserve CP.

\section{Modular invariance and the low-energy effective theory} 

The well-known target-space SL$(2,\mathbb{Z})$ symmetry, under which the spectrum and supergravity effective field theory of the heterotic string are believed to be invariant, acts on the string compactification modulus $T$ as
\begin{equation}
T \mapsto \frac{\alpha T-i\beta}{i\gamma T+\delta} \label{eq:Ttr_SL2Z}
\end{equation}
and on the charged chiral matter fields $U$ and $A$ (untwisted and twisted respectively) as
\begin{equation} 
U_i \mapsto (i\gamma T+\delta)^{-1}U_i, \qquad A_a \mapsto M_{ab} (i\gamma T+\delta)^{n_a} A_b \label{eq:matter_SL2Z}
\end{equation}
where the group element $\mathcal{M}$ of SL$(2,\mathbb{Z})$ is specified by the integers $\alpha$, $\beta$, $\gamma$ and $\delta$ satisfying $\alpha\delta-\beta\gamma=1$, the constant $n_a$ is the modular weight of the field $A_a$ and the unitary matrix $\mathbf{M}$ depends on the group element $\mathcal{M}$ but not on $T$ \cite{FerraraLT,LauerMN91}\footnote{At the one-loop level in string theory the dilaton $S$ is also shifted under the modular group, cancelling part of the anomaly due to the transformation of massless fields \cite{DerendingerFKZ,Ibanez:1992hc}.}. We use the overall modulus simplification, however the results are easily generalised to the case of several moduli. Modulus-dependent couplings of fields in the low-energy effective theory transform as a consequence of (\ref{eq:Ttr_SL2Z}) in such a way that the effective action is invariant under the combined transformations (\ref{eq:Ttr_SL2Z}), (\ref{eq:matter_SL2Z}), so modular invariance is spontaneously broken when $T$ receives a v.e.v.. Note that the modular transformation leaves the gauge charges of matter fields unchanged and, although it has a nonunitary action on the supergravity chiral fields (\ref{eq:matter_SL2Z}), the group action is unitary for canonically-normalised fields (see {\em e.g.\/}\ \cite{deCarlos:1993pd}).

We can generalise the statement of modular invariance to include any symmetry under which moduli $\hat{T}$, defined for the purpose of our discussion as fields that receive a v.e.v.\ which determines the values of low-energy coupling constants, and observable matter fields $\Phi$, defined as those whose excitations can be produced and detected experimentally, transform. The statement of invariance under the symmetry is then
\begin{equation}
\Gamma(\mathcal{M}(\hat{T}),\mathcal{M}(\Phi)) = \Gamma(\hat{T},\Phi) 
\label{eq:modinv}
\end{equation}
where $\Gamma(\hat{T},\Phi)$ is the field theory effective action functional and $\mathcal{M}(\hat{T})$ and $\mathcal{M}(\Phi)$ represent the action of the symmetry on moduli and observable fields respectively, in an obvious notation\footnote{We assume that the symmetry is unbroken above the energy scale at which the moduli v.e.v's are determined.}.

Note that when the action of modular transformations mixes different twisted sector fields, neither the individual Yukawa couplings nor the trilinear soft breaking terms in the twisted sectors of heterotic orbifolds are modular invariant \cite{Bailin:1998iz+97,mythesis}, despite the overall invariance of the effective action. The theory appears to make {\em different}\/ predictions for v.e.v.'s of $T$ that are related by a modular transformation, and which should describe the same physics. The immediate reason for this behaviour is the {\em non-Abelian}\/ modular transformation $\mathbf{M}$ of the twisted fields \cite{FerraraLT,LauerMN91}, which inevitably mixes one Yukawa coupling, {\em etc.}\/, with another.

While our results will not depend on it, we can speculate on the solution to this puzzle. It may lie in the fact that the basis of twisted states cannot be directly related to the observed mass eigenstates. For example, if two or more (left or right chiral) fermion fields are mixed by the modular transformation, then the resulting mass eigenstates must be degenerate \cite{mythesis}\footnote{Except in pathological cases where the modular symmetry permutes two mass eigenstates and in addition takes the (modulus-dependent) mass, lifetime, branching fractions, {\em etc.\/}\ of one into those of the other, and vice versa!}, which is inconsistent with experiment. The observable fields may, however, be linear combinations of the twisted fields with {\em modulus-dependent}\/ coefficients, which are invariant under the combined transformations (\ref{eq:Ttr_SL2Z}) and (\ref{eq:matter_SL2Z}). The low-energy coupling constants of the theory, written in terms of the new fields, are then modular invariant functions of $T$, even after the v.e.v.\ of $T$ is fixed.

\section{GCP transformations} 
As is well known \cite{BernabeuBG,BotellaS}, a CP transformation can include phase factors and unitary matrices acting on the observable fields. A {\em general CP transformation}\/ (GCP) is defined by the action of a group element $G$ of the internal symmetry group $\{G\}$, followed by the usual CP transformation which takes a charged scalar field to its complex conjugate, a Dirac fermion to the Dirac $\mathbf{C}$ matrix multiplying its complex conjugate, and so on. In general $\{G\}$ will include all global transformations that leave the form of the gauge couplings and kinetic terms of matter fields unchanged. In a very general notation the action of a GCP transformation with a particular group element $G$ acting on the observable fields is written as $\mathcal{GCP}[G](\Phi) = \left(G(\Phi)\right)^{\rm CP}$, where the superscript CP denotes the standard CP transformations for Weyl spinors $\psi_L$ and $\psi_R$, complex scalars $\phi$ and vectors $V^\mu$
\begin{eqnarray} &\psi_L \mapsto i\sigma^2\psi_L^*,\qquad \psi_R \mapsto -i\sigma^2\psi_R^*,& \nonumber \\ &\phi \mapsto \phi^*,\qquad V_\mu\mapsto -V^\mu.& \label{eq:nCP}
\end{eqnarray}
Then CP is conserved if and only if there is at least one element $G^x$ such that $\mathcal{GCP}[G^x]$ leaves the effective action unchanged:
\begin{equation}
\mbox{CP is conserved}\Leftrightarrow\exists\,G^x: \Gamma(\hat{T},\mathcal{GCP}[G^x](\Phi)) = \Gamma(\hat{T},\Phi). \label{eq:CPcons}
\end{equation}
This is equivalent to stating that some basis change of the observable fields will put the action into a form invariant under the standard CP transformations (\ref{eq:nCP}).

Since CP, as a gauge symmetry, cannot be explicitly violated in the heterotic string, a CP transformation acting on the values of the moduli and on the observable fields leaves the action unchanged. As mentioned above, reversing the orientation of three compactified directions is one ingredient in the construction of CP as a gauge symmetry, which implies that $T \mapsto T^*$ is the appropriate CP transformation for the $T$-modulus\footnote{The internal symmetry group does {\em not}\/ include rephasings of $T$, so we do not have the freedom to write down alternative CP transformations.}. In general the transformation of the moduli $\hat{T}$, written as $\hat{T} \mapsto \hat{T}^{\rm CP}$, must be constructed such that CP is indeed a gauge symmetry. Then we have 
\begin{equation}
\Gamma(\hat{T}^{\rm CP}, \mathcal{GCP}[G^a](\Phi)) = \Gamma(\hat{T},\Phi) \label{eq:CPspont}
\end{equation}
for at least one $G^a$.

If the group element $G$ is anomalous, the above statements for the effective action $\Gamma$ appear not to be strictly valid, since $G$ is then not a symmetry of the quantum field theory. Even if the perturbative couplings are GCP invariant, a term in the effective action proportional to Tr$\,F\tilde{F}$ for the SU$(3)_{\rm C}$ gauge group is generated by the action of $G$, so the renormalised $\bar{\theta}_{\rm QCD}$ parameter will be shifted in addition to changing sign under CP. This however contradicts our belief that GCP transformations should be physically equivalent to the ``standard'' CP transformations (\ref{eq:nCP}), which was the reason for introducing the formalism. The way out is to cancel the anomalous contributions by using some mechanism which also solves the strong CP problem: either by shifting the axion under $G$, or by axial rotations of some heavy fermions in a Nelson-Barr-like model, so that the amended action of $G$ takes $\bar{\theta}$ to $-\bar{\theta}$.

\section{CP conservation for moduli values dual to their CP transforms}
Now let us suppose that for some class $\{ \hat{T}\}$ of values for the moduli $M$ we have 
\begin{equation}
	\hat{T}^{\rm CP} = \mathcal{M}^{c}(\hat{T}) \label{eq:modcond}
\end{equation}
for some modular transformation $\mathcal{M}^c$. We would intuitively expect that this class of moduli v.e.v's should result in low-energy couplings that conserve CP, since the physics is unchanged by a modular transformation, and for these values of the moduli this is identical to a CP transformation. We now show that this rather sketchy reasoning can be justified, with one condition (to be explained) on the modular transformation of the observable fields.

We start with the statement of modular invariance (\ref{eq:modinv}) for the group element $\mathcal{M}^c$, in which we also substitute for $\mathcal{M}^{c}(\hat{T})$: then we have 
\begin{equation}
\Gamma(\hat{T}^{\rm CP},\mathcal{M}^{c}(\Phi)) = \Gamma(\hat{T},\Phi)
\end{equation}
for the class $\{ \hat{T}\}$ of values of the moduli.
Performing a (standard) CP transformation on both sides we have
\begin{equation}
\Gamma(\hat{T},(\mathcal{M}^{c}(\Phi))^{\rm CP}) = \Gamma(\hat{T}^{\rm CP},\Phi^{\rm CP})
\end{equation}
and redefining the observable fields by the group element $G^a$ such that $\Phi=G^a\tilde{\Phi}$ this becomes
\begin{equation}
\Gamma(\hat{T},(\mathcal{M}^{c}(G^a\tilde{\Phi}))^{\rm CP}) = \Gamma(\hat{T}^{\rm CP},(G^a\tilde{\Phi})^{\rm CP}) = \Gamma(\hat{T},\tilde{\Phi})
\end{equation}
where the last equality follows from (\ref{eq:CPspont}). This result should then be compared with the statement that CP is conserved (\ref{eq:CPcons}).

We deduce that, for the set of values of moduli satisfying (\ref{eq:modcond}), CP is conserved, given only that the action of $\mathcal{M}^{c}$ on the observable fields is an element of the internal symmetry group $\{G\}$. But for canonically-normalized fields $\Phi$, the modular symmetry acts by a unitary transformation that leaves invariant the kinetic terms and gauge quantum numbers of the fields, so it should be allowed as an element of $\{G\}$. Hence CP is conserved in the low-energy field theory for the class $\{\hat{T}\}$ of moduli v.e.v.'s. We assume throughout that $\bar{\theta}_{\rm QCD}$ is zero, in which case it remains zero after GCP and non-anomalous global transformations as explained earlier.

\section{Applications} 
We can immediately specialise to the case when the modular invariance group is SL$(2,\mathbb{Z})$ and the $T$ moduli are the only source of CP violation considered. Then by considering the generators $T\mapsto 1/T$, $T\mapsto T+i$ we see that CP is conserved for $T$ on the unit circle and on the lines Im$\,T=\pm 1/2$, that is, precisely on the boundary of the fundamental domain $\mathcal{F}$.
Thus Giedt's claim \cite{Giedt2000} that the complex phases of Yukawa couplings for the v.e.v.'s $T=e^{\pm i\pi/6}$ in a particular string orbifold do not produce CP violation, follows immediately as a special case. 

So if the observed CP violation is due to the v.e.v.\ of $T$, the modulus must lie inside the fundamental domain: this is difficult to achieve for stabilization mechanisms which respect the modular symmetry \cite{CveticFILQ} but possible using gaugino condensation with universal threshold corrections \cite[and references therein]{BailinKL2000}. It will then depend on a particular model as to how CP will be violated in the low-energy couplings. The possibility that the ``supersymmetric CP problem'' could be solved in heterotic string models by complex $T$ values lying on the boundary of $\mathcal{F}$ \cite{IbanezLust}, leading to soft terms that vanish or appear to conserve CP, is ruled out as a model of CP violation, since the Yukawa couplings would then also conserve CP. Conversely, the possibility that soft supersymmetry-breaking terms are the only source of CP violation \cite{AbelF,Brhlik2000} is not motivated either in this scenario (although the possibility that the CKM phase could vanish for $\langle T\rangle$ at isolated points within $\mathcal{F}$ remains). We speculate that for generic modulus v.e.v.'s CP will be violated in all possible sets of couplings, which appears to favour the supersymmetric scenarios of approximate CP (compare \cite{Eyal:1998bk}) or of large soft phases which evade the electric dipole moment bounds by cancellations or by a particular flavour structure \cite{Ibrahim98,BrhlikGK99,Pokorski:2000hz}. Approximate CP might be realized for modulus v.e.v.'s close to the CP-conserving boundary of $\mathcal{F}$.

One might object that the shift of the dilaton
\[ S\mapsto S-\frac{3\delta_{\rm GS}}{(8\pi^2)} \ln (i\gamma T+\delta) \]
under modular transformations, mentioned earlier, is inconsistent with the CP transformation (\ref{eq:modcond}), where $\hat{T}$ should include $S$. Our reasoning is then only strictly valid in the presence of an axionic symmetry which renders the value of Im$\,S$ unobservable \cite{IbanezLust,Choi97,GeorgiKN}, or for particular values of Im$\,S$ satisfying (\ref{eq:modcond}). Note that for Im$\,T=\pm 1/2$ the CP transformation $T\mapsto T\mp i$ is realized by a modular transformation with $\ln(i\gamma T+\delta)=0$, while for $T$ on the unit circle the modular transformation $\mathcal{M}^c$ has $\delta=0$ and $\ln(i\gamma T+\delta)$ pure imaginary. In fact the soft terms in heterotic string models seem not to depend on Im$\,S$, if supersymmetry-breaking is dominated by gaugino condensation in a single gauge group (see {\em e.g.\/}\ \cite{Bailin:1998iz+97}). But in general, the presence of more than one term in the superpotential of the effective theory will lead to a strong dependence on Im$\,S$ \cite{Choi97}, and CP may be violated, even for $T$ on the boundary of $\mathcal{F}$.

To get a clearer picture of this scenario, consider the GCP transformation $\mathcal{GCP}[G^{a}\mathcal{M}^c]$ acting on $\Gamma(T,S,\Phi)$. We have 
\begin{equation}
	\Gamma(T,S,\Phi)\stackrel{\mathcal{GCP}[G^{a}\mathcal{M}^c]}{\longmapsto} \Gamma^{\rm GCP} \equiv \Gamma(T,S,G^{a*}(\mathcal{M}^{c*}(\Phi^{\rm CP}))).
\end{equation}
Then using the statement of spontaneous CP violation (\ref{eq:CPspont}) which now takes the form
\[ \Gamma(T^*,S^*,(G^{a}\Phi)^{\rm CP}) = \Gamma(T,S,\Phi) \]
(note that $G^{a*}$ must be a symmetric unitary matrix satisfying $G^{a*}G^a=\mathbf{1}$) this becomes
\begin{eqnarray}
	\Gamma^{\rm GCP} &=& \Gamma(T^*,S^*,\mathcal{M}^{c}(\Phi)) \\
	&=& \Gamma(T,S^*-3\Delta_{\rm GS}\ln(i\gamma'T^*+\delta'),\Phi)
	\label{eq:GammaGCP3}
\end{eqnarray}
where $\Delta_{\rm GS}=\delta_{\rm GS}/(8\pi^2)$ and the last equality follows from modular invariance under the group element $(\mathcal{M}^{c})^{-1}$, specified by integers $\alpha'$, $\beta'$, $\gamma'$, $\delta'$, for values of $T$ on the boundary of $\mathcal{F}$. Then for these values of $T$ {\em the only change in the theory under the GCP transformation is the change in Im}\/$\,S$. 

If the CP invariance condition (\ref{eq:modcond}) holds then we have
\[ S^*=S-3\Delta_{\rm GS}\ln(i\gamma T+\delta) \] 
so the expression $S^*-3\Delta_{\rm GS}\ln(i\gamma'T^*+\delta')$ reduces to $S-3\Delta_{\rm GS}\ln(-\gamma\gamma')$ for values of $T$ on the unit circle for which we choose $\mathcal{M}^c$ to take $T$ to $1/T$. For $(\mathcal{M}^c)^{-1}$ to reverse the action of $\mathcal{M}^c$ (as it should) we require $\gamma\gamma'=-1$, so as we expected $\Gamma^{\rm GCP}=\Gamma$ by eqn.\ (\ref{eq:GammaGCP3}) and CP is conserved. But the CP conservation condition now holds only for particular values of Im$\,S$, which we would not expect to be picked out by any stabilization mechanism. This opens up the intriguing possibility that CP could be violated {\em via the $S$-dependence of the low-energy couplings only}, if $\langle T\rangle$ is on the unit circle and no modular transformation can satisfy (\ref{eq:modcond}) for $T$ without requiring a shift in $S$. Of course for the other CP-conserving values Im$\,T=0$, $\pm 1/2$, a nonzero imaginary part for $S$ could break CP in the same way. Thus the Yukawa couplings, which in the heterotic string appear not to be functions of $S$, would on their own conserve CP, while soft supersymmetry-breaking terms (for example) might break it via a complex $F^S$. This would be a highly constrained scenario which might result from known supersymmetry-breaking mechanisms, for example multiple gaugino condensation, and could easily be tested.
   
The results just obtained for the SL$(2,\mathbb{Z})$ invariance of the $T$ modulus can be generalised to other invariance groups which are spontaneously broken by potentially CP-violating v.e.v.'s. In the simplest models of spontaneous CP violation \cite{Lee_Weinberg}, to give a trivial example, one may impose $\mathbb{Z}_N$ symmetries (not to be explicitly broken) acting on the Higgses, in which case CP is unbroken for vacua which are connected to their CP conjugates by a $\mathbb{Z}_N$ transformation.

We might hope that our result could be extended to models based on fundamental theory beyond the perturbative heterotic string; however, more work is needed to find the appropriate action of CP in the full theory and the quantum symmetries acting on CP-violating quantities. The obvious extension is to ``heterotic M-theory'' defined as the strong coupling limit of the heterotic string \cite{Horava_Witten}. For the standard embedding of the spin connection in the gauge group one would expect that the strong-coupling limit should be continuously connected to the perturbative string by expanding the 11th dimension (see \cite{NillesS97}), in which case CP should survive as a gauge symmetry and modular invariance should also hold. Note that explicit calculations of soft breaking terms in the ``M-theory'' limit of large $S$ and $T$ show exponentially vanishing imaginary parts \cite{BailinKL2000}, so even for v.e.v's of $T$ inside $\mathcal{F}$ the soft phases may be negligibly small (however the status of CP violation in Yukawa couplings at large T is not known).

For non-standard embeddings including fivebranes the status of modular invariance and CP violation is less clear, although recent work on the four-dimensional effective action including fivebranes \cite{DerendingerS} suggests that the {\em r\^ole}\/ of the fivebrane moduli may be similar to that of the chiral $S$ and $T$ fields, namely as complex scalars potentially contributing to spontaneous CP violation.

It is unclear if CP survives as a gauge symmetry in the presence of D-branes or warped compactifications. If not, then explicit CP violation is unavoidable: the consequences may include the need to have an axion to solve the strong CP problem and a lack of control over ``soft phases'' in supersymmetric models. Also, the possible CP-violating background fields in Type I/Type IIB constructions have yet to be completely counted. The stringy dualities are different: instead of the target-space self-duality of the heterotic string, Type I/Type IIB compactifications have T-duality symmetries which relate one model to another with different sets of branes and coupling constants (see {\em e.g.\/}\ \cite{Ibanez:1999rf}). Target-space modular {\em invariance}\/ has been conjectured for Type IIB orientifolds, by analogy with the corresponding heterotic orbifold, but one-loop calculations do not appear to support its existence as a quantum symmetry \cite{Lalak:2000ex}. 

We would expect constraints, similar to those described above, to apply to possible sources of CP violation in more general ``brane world'' models in which observable matter fields are localized in extra-dimensional space; however, a more complete theoretical description of such models would be required in order to make definite statements about CP violation (see \cite{Sakamura:2000ik} for an interesting exception).

\subsection*{Acknowledgements}
The author wishes to thank David Bailin for the original motivation of this work and for many discussions on the subject of modular invariance, and Gordy Kane, Dan Chung and Lisa Everett for comments on earlier versions of the paper. This research was supported in part by DOE Grant DE-FG02-95ER40899 Task G.

\end{document}